\documentclass[12pt,aps,amsmath,latexsym,amsfonts]{JHEP3}

\usepackage{epsfig}
\usepackage{amsfonts,amssymb,amsmath}
\usepackage[hang]{subfigure}

\newcommand{\be}{\begin{equation}}
\newcommand{\ee}{\end{equation}}
\newcommand{\bea}{\begin{eqnarray}}
\newcommand{\eea}{\end{eqnarray}}

\def\real{\mathbb R}
\def\half{\textstyle{\frac{1}{2}}}

\title{Spinflation with Angular Potentials}
\author{
Ruth Gregory\thanks{Email: r.a.w.gregory@durham.ac.uk}~~and
Dariush Kaviani\thanks{Email: dariush.kaviani@durham.ac.uk} \\
Institute for Particle Physics Phenomenology
and Centre for Particle Theory, \\
Durham University, South Road, Durham, DH1 3LE, UK}

\abstract{
We investigate in detail the cosmological consequences of realistic angular 
dependent potentials in the brane inflation scenario. 
Embedding a warped throat into a compact Calabi-Yau space with all 
moduli stabilized breaks the no-scale structure and induces 
angular dependence in the potential of the probe D3-brane. 
We solve the equations of motion from the DBI action 
in the warped deformed conifold including linearized perturbations 
around the imaginary self-dual solution.
Our numerical solutions show that angular dependence is 
a next to leading order correction to the dominant radial motion
of the brane, however, just as angular motion typically
increases the amount of inflation (spinflation), having additional
angular dependence also increases the amount of inflation.
We also derive an analytic approximation for the number
of e-foldings along the DBI trajectory in terms of the 
compactification parameters.
}

\preprint{DCPT-11/37\\ IPPP/11/40\\DCPT/11/80}

\begin{document}
\newcommand{\zed}{$\mathbb{Z}_2$}

\section{Introduction}

It is now generally accepted that the very early Universe underwent a period
of rapid expansion, known as inflation, 
resulting in a very nearly flat, homogeneous and 
isotropic initial state. While a simple scalar field model
of inflation with a suitable potential satisfies many of the
cosmological requirements, such models are rather ad hoc from the
high energy particle theory point of view. The challenge is to find a
theory which has a clear derivation from a fundamental high energy
theory incorporating gravity at the quantum level. 
String theory naturally is the prime candidate for such a
fundamental theory, but one of the major challenges has been to
show that in a non-vanishing fraction of the vast number of string vacua the
inclusion of various compactification effects in the effective field theory
leaves a suitable inflationary model.

In recent years there has been significant progress within
the `KKLT' framework, \cite{KKLT}, in which the older
idea of brane inflation \cite{brinfl}, is realised via 
the motion of a brane in internal, hidden, extra
dimensions \cite{KKLMMT} (see also \cite{followon,BDKKM,tip,tipP,IIB} 
and the reviews \cite{reviews}).
In these scenarios the inflaton potential (of the position of
a probe D3-brane) is flattened by fine-tuned
cancellation of correction terms from moduli stabilizing wrapped 
branes and bulk effects so that inflation can occur.
However, whether or not the potential can be made flat to meet the slow-roll
conditions, DBI inflation, \cite{ST}, is also possible, even when the 
potential is steep. In this case, while the motion of the brane can
be strongly relativistic (in the sense of a large $\gamma$-factor)
strong warping of the local throat region renders their contribution
to the local energy density subdominant to that of the inflaton
potential terms. 

In either case, the approach taken is to consider a D3-brane 
moving in a warped throat region of a Calabi-Yau flux compactification 
of type IIB theory with ISD conditions \cite{GKP}.
However, when the UV end of the warped throat
is attached to the compact Calabi-Yau space with all moduli stabilized,
violations of ISD conditions with important implications for the 
action of the brane are expected.
In particular, the potential of a mobile D3-brane in the compactified throat
geometry receives angular dependent corrections, \cite{BDKKM},
which until recently, have been largely neglected (although
see \cite{tip,tipP,ABMX}). In slow roll
inflation, it was presumed that the angular directions would
stabilise rapidly, with the radial (slow-roll) direction dominating
the inflationary trajectory. However, for generic brane motion 
the effect of angular motion is less clear, particularly in the presence
of angular terms in the potential. 

Angular motion of branes was initially explored in the probe limit, i.e.\
where the brane does not back-react at all on either the internal or
external dimensions. Unsurprisingly, the angular motion has a
conserved momentum, which can give interesting brane universes 
(see e.g.\ \cite{braneon,EGTZ, sling}), however the mirage style, 
\cite{mirage}, interpretation of the cosmology of these universes 
leads to a rather unsatisfactory picture. Based on the probe understanding,
it was conjectured that angular motion would not affect a more realistic
inflationary scenario to any great extent, an expectation largely borne
out by the ``spinflation'' study, \cite{spin}, which found a marginal
increase of a couple of e-foldings due to angular motion, coming mainly
from the initial stages of inflation before the angular momentum
becomes redshifted away. This increase is however parameter
sensitive, a point not noted in this original study.
In \cite{spin}, a general DBI-inflationary universe was considered
near the tip of the warped deformed conifold throat, the 
Klebanov-Strassler (KS) solution \cite{KS}, with a simple
radial brane potential; with a more realistic potential including angular
terms, the spinflationary picture could potentially be rather different.

In this paper therefore, we investigate the cosmological implications of
including angular dependence in the DBI brane inflation scenario. 
Building on the results of Baumann \emph{et al.} \cite{BDKKM}, 
we consider D3-brane
motion in the warped throat region of the compact Calabi-Yau subject to UV
deformations of the geometry that induce angular dependent corrections in the
potential of the probe D3-brane. Taking into account linearized perturbations
around the ISD solution, we solve the D3-brane equations of motion from the DBI
action with angular dependence induced by the leading correction to the
potential allowed by the symmetries of the compactification. 
Our aim is to consider angular momentum in a fully UV/IR consistent
fashion, and to account for angular momentum in a more general
potential. As with the simpler radial spinflation potential, our numerical 
solutions show that angular dependence tends to increase the inflationary 
capacity of a trajectory, increasing the number of  e-foldings, albeit at a 
subdominant level. To a large extent, the trajectories are still predominantly
radial, however, do exhibit rotational motion due to the angular potential.

An important question is how generic such trajectories are. In general, as the
brane arrives in the throat, one expects a range of initial conditions in
terms of angular values and velocities. We find that the D brane
trajectories and number of e-foldings are dependent more on
model parameters than on the initial condition of the brane motion,
thus indicating that the results of our investigation are reasonably
robust.

\section{The supergravity background}

In brane inflation, a mobile D3 brane (or anti-brane) is embedded in the
internal manifold, with its four infinite dimensions
parallel to the 4D noncompact universe.
The position of the brane on the internal manifold then
provides an effective 4D scalar field - the inflaton. 
The 10D set-up is assumed to be a flux compactification of type IIB
string thory on an orientifold of a Calabi-Yau threefold (or an F-theory
compactification on a Calabi-Yau fourfold) \cite{GKP}. We are
interested in the situation where fluxes  have generated a warped 
throat in the internal space, and will be examining primarily the
deep throat region. 

We are therefore considering backgrounds in low-energy 
IIB supergravity, which in the Einstein frame can be
represented by the action
\bea
S_{\,\text{IIB}} & = & - \frac{1}{2\kappa_{10}^2}\int d^{10}x \sqrt{|g|}
\left[ \mathcal{R} - \frac{|\,\partial\tau|^2}{2(\text{Im}\tau)^2} - \frac{
|\,G_{3}|^2}{12\,\text{Im}\tau} - \frac{|\,\tilde{F_{5}}|^2}{4\cdot5\,!}\right]
\nonumber \\ 
&&+\frac{1}{8i\kappa_{10}^2}\int \frac{C_{4}\wedge G_{3} \wedge
{\bar G}_3}{\text{Im}(\tau)}+S_{\,\text{loc}},
\label{2BEA}
\eea
where $\tau=C_{0}+ie^{-\phi}$ is the axion-dilaton field,
$G_3=F_{3}-\tau H_{3}$ is a combination of the RR and NSNS
three-form fluxes $F_{3}=dC_{2}$ and $H_{3}=dB_{2}$, and
\be
{\tilde F}_5 = d C_4 - \half C_2 \wedge H_3 + \half B_2 \wedge F_3
\qquad (+ \; {\text{dual}})
\ee
is the 5-form field, whose self duality must be imposed by hand.
The constant $\kappa_{10}^2$ is the 10D gravitational coupling:
\be
\kappa_{10}^2 = M_{10}^{-2} = \frac{(2\pi)^7 g_s^2 {\alpha^{\prime}}^4}{2} \;.
\label{kappa10}
\ee

In a flux compactification, we are assuming a block diagonal Ansatz
for the metric:
\be
\label{10Dmetric}
ds_{10}^2=e^{2A(y)}{g_{\mu\nu}dx^{\mu}dx^{\nu}}
- e^{-2A(y)}{{\tilde g}_{mn}dy^{m}dy^{n}},
\ee
in which the warp factor, $e^{4A(y)}$, depends only on the internal
coordinates $y^m$, the internal metric ${\tilde g}_{mn}$ is independent
of the spacetime coordinates (and will be taken to be a known supergravity
solution) and the 4D metric, $g_{\mu\nu}$, is taken as Minkowski for
the computation of the supergravity flux background, but ultimately
will be assumed to have an FRW form
once the general cosmological solution is sought.

Following \cite{GKP}, we take the self-dual 5-form to be given by
\be
\label{SD5}
\tilde{F}_{5}=(1+\star_{10})\Big[d\alpha(y)\wedge dx^{0}\wedge dx^{1}\wedge
dx^{2}\wedge dx^{3}\Big],
\ee
in the Poincar\'e invariant case, where $\alpha(y)$ is a function
of the internal coordinates. The Einstein equations and 5-form 
Bianchi identity then imply
\be
\label{trace}
{\widetilde \bigtriangleup} \Phi_{\pm} = \frac{e^{8A+\phi}}{24}|\,G_{\pm}|^2
+e^{-4A}|\,\nabla\Phi_{\pm}|+\text {local},
\ee
where ${\widetilde\bigtriangleup}$ is the Laplacian with respect to the
6D metric ${\tilde g}_{mn}$, and 
\be
G_{\pm}\equiv(i\pm\star_{6})G_{3}, \;\;\;\;\;\;\ \Phi_{\pm}\equiv
e^{4A}\pm\alpha.
\ee
For $G_{-}= 0$, i.e.\ $\star_6 G_3 = iG_3$, the flux $G_3$ is imaginary
self-dual (ISD), and the background metric ${\tilde g}_{mn}$ is a 
Calabi-Yau metric with the 5-form flux given by $\alpha = e^{4A}$ \cite{GKP}.

The particular concrete example we will be interested in is where the 
background is the warped deformed conifold, or Klebanov-Strassler (KS),
solution \cite{KS}. Here, the internal metric is a strongly warped and
deformed throat, which interpolates between a regular $\real^3 \times S^3$
tip, to an $\real \times T^{1,1}$ cone in the UV. The metric
is usually presented as: 
\bea
{\tilde g}_{mn} dy^m dy^n &=&
\frac{\epsilon^{4/3}}{2} K(\eta)\Big[
\frac{1}{3K(\eta)^3}\{d\eta^2+(g^{5})^2\}
+\cosh^2\frac{\eta}{2}\{(g^{3})^2+(g^ {4})^2\}\nonumber \\
&& \hskip 2cm + \sinh^2\frac{\eta}{2}\{(g^{1})^2+(g^{2})^2\}\Big],
\label{KS6}
\eea
where $\epsilon$ is the deformation
parameter of the conifold; it is a dimensionful parameter 
($[\epsilon^{2/3}]=L$), and sets a scale for the throat as 
we will see presently. The `radial' coordinate, $\eta$, is 
chosen so that the function $K$ is expressible
explicitly in analytic form:
\be
\label{KSfn}
K(\eta)=\frac{(\sinh\eta\cosh\eta-\eta)^{1/3}}{\sinh\eta}\; ,
\ee 
and the $g^i$'s are forms representing the angular directions, 
given explicitly by
\be
\label{gdef}
g^{1,3}=\frac{e^{1}\mp e^{3}}{\sqrt{2}},\;\;\;\ g^{2,4}=\frac{e^{2}\mp
e^{4}}{\sqrt{2}},\;\;\;\ g^{5}=e^{5}
\ee
with
\bea                                                                  
\label{edef}
e^{1} & = & -\sin\theta_{1} d\varphi_{1},\;\;\;\;\;\;\  e^{2}= d\theta_{1},
\;\;\;\;\;\;\  e^{3}= \cos\psi \sin\theta_{2}d\varphi_{2}-\sin\psi d\theta_{2},
\nonumber \\
e^{4} &= & \sin\psi\sin\theta_{2}d\varphi_{2}+\cos\psi d\theta_{2}, 
\;\;\;\;\;\;\
e^{5}= d\psi+\cos\theta_{1}d\varphi_{1}+\cos\theta_{2}d\varphi_{2}.
\eea
(For details on the warped deformed conifold, and coordinate
systems, see e.g.\ \cite{coni}.)

It is also useful to visualise this metric in terms of a proper
radial coordinate
\be
r(\eta) = \frac{\epsilon^{2/3}}{\sqrt{6}} \;
\int_0^\eta \frac{dx}{K(x)}
\label{rproper}
\ee
which measures the actual distance up the throat in the six-dimensional
metric $\tilde g$. The metric can now be written as:
\bea
d{\tilde s}_6^2 &=& dr^2 + r^2 \left [ \frac{C_3^2(r)}{9} (g^{5})^2
+ \frac{C_1^2(r)}{6} \{(g^{3})^2+(g^ {4})^2\}
+ \frac{C_2^2(r)}{6} \{(g^{1})^2+(g^{2})^2\} \right ] \;\;\;\;\;
\label{KS6r}
\eea
where the functions $C_i(r)$ are given implicitly from (\ref{KS6}), and
are shown in figure \ref{fig:KSrfns}. At small $r$: 
\be
r\sim \frac{\epsilon^{2/3}}{3^{1/6} 2^{5/6}} \; \eta
\;\;\;\;,\;\;\;\;\; K \simeq \left (\frac23\right)^{1/3}
\ee
and the metric (\ref{KS6r}) smoothly closes off at $r=0$ with a
finite $S^3$ of radius $\epsilon^{2/3}/12^{1/6}$. At large $r$,
or for $\eta \sim 10-15$, the $C_i$'s become unity, and the throat 
explicitly takes the form of a cone: $\real\times T^{1,1}$.
Thus, $\epsilon^{2/3}$ gives the radius of the nonsingular 
$S^3$ at the base of the throat, and the scale at which the throat
asymptotes the $T^{1,1}$ cone: It sets the IR scale of the geometry.
\FIGURE[ht]{
\epsfig{file=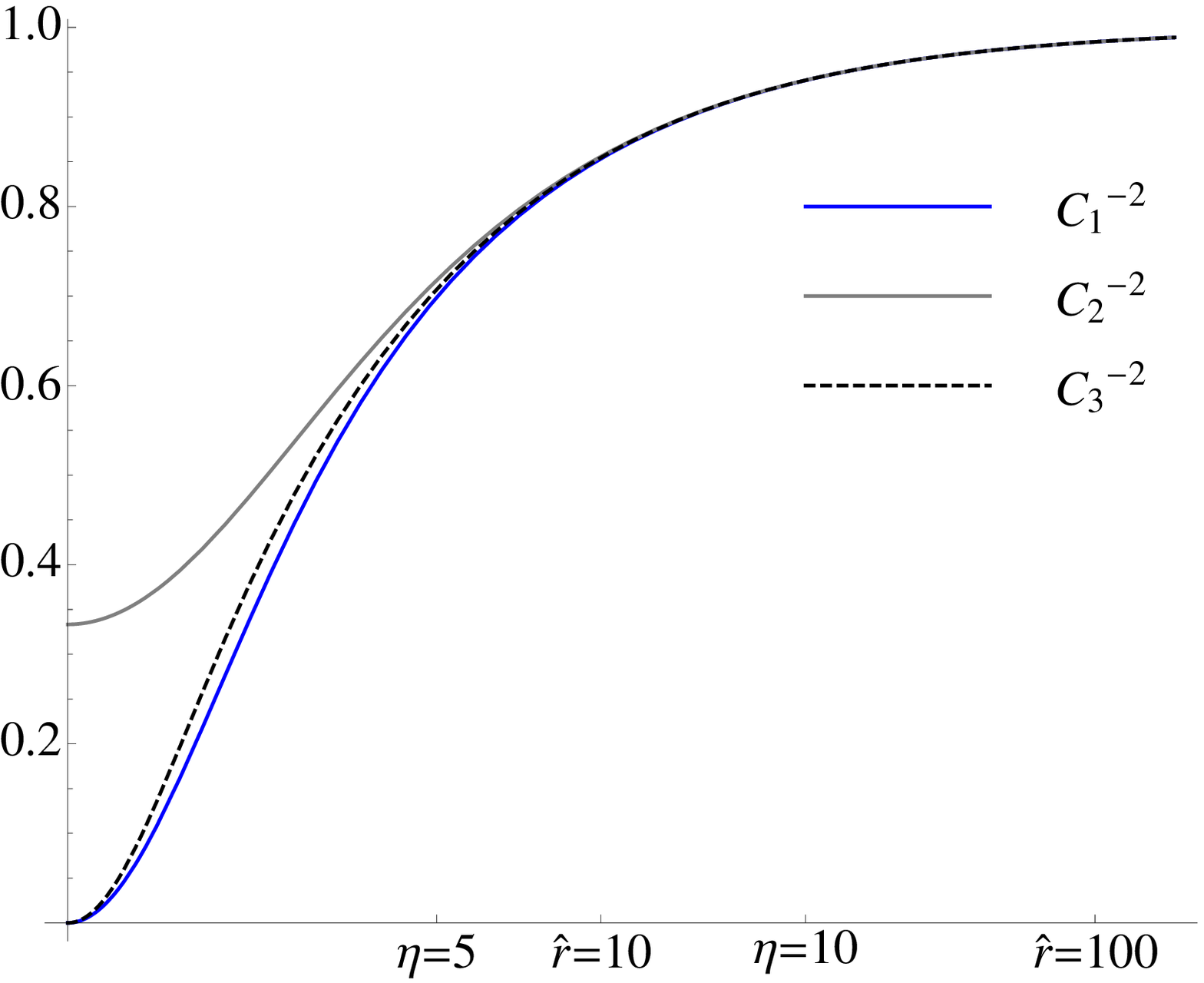,width=.7\textwidth}
\caption{A plot of the metric functions $C_i$, which shows how the
deformation of the conifold increases as the tip is approached.
The metric asymptotes the $T^{1,1}$ cone, but near the origin,
$rC_1$ and $rC_3$ remain finite leaving the nonsingular $S^3$ 
metric at the tip of the `cone'. The axis is labelled both in terms
of the original $\eta$ coordinate, as well as the normalized proper
distance up the throat, ${\hat r} = \epsilon ^{-2/3}r$.}
\label{fig:KSrfns}
}

In terms of the $\eta$ coordinate, the warp factor has the form, \cite{KS}
\be
\label{KSh}
e^{-4A} = 2 (g_{s}M\alpha^{\prime})^{2}\; \epsilon^{-8/3} \;  I(\eta),
\ee
where\footnote{Note that this definition of $I$ relocates a factor
of $2^{1/3}$ relative to the KS expression.}
\be
\label{KSint}
I(\eta)\equiv \int_{\eta}^{\infty}{dx\frac{x\coth x-1}{\sinh^{2}x}}
(\sinh x \cosh x -x)^{1/3}.
\ee
Thus for small $\eta$
\bea
e^{-4A} &\sim& 2 (g_{s}M\alpha^{\prime})^{2}\; \epsilon^{-8/3} \;
\left [ I(0) - \frac{\eta^2}{12} \left (\frac23 \right)^{1/3}\right]\\
&=& 2 (g_{s}M\alpha^{\prime})^{2}\; \epsilon^{-8/3} \;
\left [0.5699 - \epsilon^{-4/3} \frac{r^2}{3} \right ] \nonumber
\eea
and for large $\eta$:
\bea
e^{-4A} &\sim& 3 \cdot 2^{1/3} (g_{s}M\alpha^{\prime})^{2}\; \epsilon^{-8/3} \;
(\eta - 1/4) e^{-4\eta/3} \\
&=& \frac{27}{8} \; \frac{(g_{s}M\alpha^{\prime})^2}
{r^4} \left ( \ln\frac{r^3}{\epsilon^2} + \ln\frac{4\sqrt{2}}{3\sqrt{3}} 
-\frac14 
\right )\nonumber
\eea
which shows how the warp factor interpolates between the 
Klebanov-Tseytlin (KT) form, \cite{KT},
for large $r$, to a smooth cap for $r\epsilon^{-2/3} \lesssim 1$. 
In addition, for sufficiently small $\epsilon$, there is 
an intermediate adS region for $| \ln r| \ll |\ln \epsilon^{2/3}|$, 
in which the logarithmic dependence on $r$ is subdominant
to the deformation term.

In an ISD compactification such as the warped deformed conifold, only the
complex-structure moduli are stabilized but not the K\"ahler moduli (that
characterize the size of the internal Calabi-Yau). When the no-scale structure
is broken to stabilize the K\"ahler moduli, perturbations around 
the ISD solution are expected. In supergravity, these perturbations are
sourced by attaching the UV end of the throat to a fully stabilised Calabi-Yau
sector. Recall that in an ISD solution, we have $G_-=0$ and $\alpha = e^{4A}$,
so that $\Phi_-=0$. In a perturbed solution, $\Phi_-$ is nonvanishing and its
value is determined by the solution of the supergravity equation of motion
(\ref{trace}). The dominant source for $\Phi_-$ is the IASD flux, $G_-$,
however this flux sources only second order perturbations, hence 
the Einstein equations and five-form 
Bianchi identity imply that the perturbations of $\Phi_{-}$ 
around ISD conditions satisfy, at linear level, the 6D Laplace 
equation with respect to the unperturbed metric \cite{BDKKM,GKP}:
\begin{equation}
\label{ISDLap}
{\tilde \Delta}\Phi_{-}=0.
\end{equation}
In a noncompact throat geometry, we can solve the Laplace equation 
and obtain the structure of $\Phi_-$, which then feeds in 
to a potential for the D3-brane motion. (See \cite{BDKKM}
for a detailed explanation and computation of potentials away from
the tip of the throat.)

In most inflationary models, the potential is computed away
from the tip of the throat, and the geometry
is approximated by $AdS_{5}\times T^{1,1}$. In this case, the
angular part of the Laplace equation, (\ref{ISDLap}) is relatively
straightforward, and solutions take a particularly clean form:
\be
\label{KTsoln}
\Phi_-(r,\Psi)=\sum_{L,\,M}\Phi_{LM}
\Bigg(\frac{r}{r_{\text{UV}}}\Bigg)^{\Delta(L)}Y_{LM}(\Psi)\,,
\ee
where $r$ is the proper radial distance in the metric ${\tilde g}_{mn}$,
(\ref{rproper}), and
\be
\label{eval}
\Delta(L) \equiv -2+\sqrt{6[\,l_{1}(l_{1}+1)+l_{2}(l_{2}+1)-R^2/8]+4}
\ee
is the radial eigenfunction weight, coming from the eigenvalues of the angular
eigenfunctions, $Y_{LM}(\Psi)$, of the Laplacian on $T^{1,1}$, \cite{CDA}.
$L\equiv (l_{1},l_{2},R)$, $M\equiv (m_{1},m_{2})$ 
label the $SU(2)\times SU(2)\times U(1)_{R}$ quantum numbers
under the corresponding isometries of $T^{1,1}$, and $\Phi_{LM}$ are constant
coefficients. The leading order terms of interest have the
lowest $\Delta(L)$, the smallest eigenvalues corresponding to non-trivial
perturbations being
\bea                                                                 
\Delta &=&\frac{3}{2}\;\;\;\;\;\;\ \text{for}\;\;\ (l_{1}, l_{2},
R)=(\half,\half,1), \;\;\; Y_{LM} \sim \cos\textstyle{\frac{\theta_1}{2}}
\cos\textstyle{\frac{\theta_2}{2}} e^{i(\phi_1+\phi_2+\psi)/2} \label{del32}\\
\Delta &=& 2\;\;\;\;\;\;\;\ \text{for}\;\;\ (l_{1}, l_{2}, R)=(1,0,0),(0,1,0)
\;\;\;\; Y_{LM} \sim \cos \theta_{i} . \label{del2}
\eea
In \cite{BDKKM}, the first mode, (\ref{del32}), was used to construct
an inflection potential for the inflationary universe, however, this mode
is not allowed in the warped deformed conifold, as the $U(1)_R$ isometry 
is broken to a discrete $\mathbb{Z}_{2}$ and therefore (\ref{del32}) is 
forbidden. 

The Laplacian on a general warped deformed conifold was computed in
\cite{KK,PKKL}, although the radial eigenfunctions were not computed
explicitly. Fortunately, since we are only interested in the low
lying states dependent on only one angle, $\theta_i$,
the Laplacian reduces to\footnote{We agree with the correction
noted by Pufu et al., \cite{PKKL}, to the Laplacian in \cite{KK}.}
\be
\frac{1}{\sinh^2\eta} \left [ 6\frac{\partial\;}{\partial\eta}
K^2(\eta)\sinh^2\eta \frac{\partial\;}{\partial\eta}
+\frac{4}{K(\eta)} \frac{\cosh\eta}{\sin\theta_i}
\frac{\partial\;}{\partial\theta_i} \sin\theta_i 
\frac{\partial\;}{\partial\theta_i} \right ] \Phi_- =0 
\ee
which for $\ell=1$ can be solved analytically
to obtain the eigenfunction:
\be
\Phi_-(\eta,\Psi) \propto (\cosh\eta\sinh\eta - \eta)^{1/3} \cos \vartheta
\label{KSefn}
\ee
where $\vartheta$ stands for {\it either} $\theta_1$ or $\theta_2$.
Since $r\propto e^{\eta/3}$ for large $\eta$, it is easy to see that this
corresponds to the second eigenfunction, (\ref{del2}), of the mid-throat
region. Note however that there is no clean expression of this 
eigenfunction in terms of the radial coordinate

We will thus use this leading order correction to the background
near the tip of the throat to investigate the effect of angular 
perturbations on the brane motion in inflation.

\section{Brane inflation with an angular potential}

The premise of brane inflation is that a D3-brane, extended in the
non-compact dimensions, can move around on the internal manifold
in such a way that the `scalar' field $y^m$, representing the location
of the brane on the internal manifold, plays the role of the inflaton. 
For a mobile D3-brane moving on a supergravity background, the effective
action is given by combining the DBI effective action for the worldbrane
coordinates, and the Wess-Zumino coupling to the RR-background. Choosing
a gauge which aligns with the coordinate system 
($X_{D3}^a = (x^\mu, y^m(x^\mu)$), gives the explicit form:
\bea
S_{DBI} +S_{WZ} &=& -T_3\, \int{d^4\xi\, \sqrt{-\text{det}
(\gamma_{a b}+ {\mathcal F}_{a b} )}}\,+
\,T_3 \,\int_{\mathcal W}{C_4}\,,\nonumber \\
&=&  -T_3 \int d^4 x \sqrt{-g} \left [ e^{4A} \sqrt{\text{det}(\delta^\mu_\nu
- e^{-4A} y^{m,\mu} y^n_{,\nu} {\tilde g}_{mn} )} - \alpha \right ]\,,
\label{D3act}
\eea
where $T_3 = 1/(2\pi)^3 g_s \alpha^{\prime2}$ is the D3-brane tension.
The energy momentum from this action,
\be
T_{\mu\nu} = T_3 \left \{ e^{4A} \sqrt{\text{det}(\delta^\mu_\nu
- e^{-4A} y^{m,\mu} y^n_{,\nu} {\tilde g}_{mn} )} \left (
g_{\mu\nu} - e^{-4A} y^m_{,\mu} y^n_{,\nu} {\tilde g}_{mn} \right)
- \alpha g_{\mu\nu} \right \}
\label{braneem}
\ee
can then drive gravitational physics in the noncompact dimensions.
In general, there will also be additional terms coming from corrections
to the supergravity background which will appear as effective potential
terms for the internal coordinates.

The 4D effective gravitational action can be obtained by integrating
out (\ref{2BEA}) for the background (\ref{10Dmetric}):
\be
S = -\frac{1}{2\kappa_{10}^2} \int d^{10} x \sqrt{g_{10}} \; {\cal R}
\to -\frac{1}{2\kappa_{10}^2} \left ( \int d^{\,6} y \; e^{-4A} \sqrt{\tilde g}
\right ) \int d^4 x \sqrt{g} R (g) \;\;,
\ee
which gives the 4D Planck mass as
\be
M_p^2 = \frac{1}{\kappa_{10}^2}\int e^{-4A} \sqrt{\tilde g} \; d^{\,6} y
\geq \frac{1}{6g_s^2\alpha^{\prime4}} \left ( \frac{\epsilon}{2\pi}
\right )^4 \int _0^{\eta_{UV}} e^{-4A} \sinh^2 \eta \; d\eta \;\;,
\label{Mpbd}
\ee
and provides a constraint on the parameters of the solution, \cite{BMbd}.

For an inflationary solution, we will take the position of the D3
brane to be homogeneous, i.e.\ $y^m = y^m(t)$, and we will assume
an FRW 4D metric:
\be
g_{\mu\nu} dx^\mu dx^\nu = dt^2 - a^2(t) d{\bf x}^2 \;.
\ee
Defining the relativistic $\gamma-$factor as
\be
\gamma = 1/\sqrt{1 - e^{-4A} {\dot y}^m {\dot y}^n {\tilde g}_{mn}}\;,
\label{gammadef}
\ee
the energy-momentum tensor of the brane is:
\bea
T^\mu_\nu = T_3 \, {\rm diag} \left ( e^{4A} \gamma -\alpha,
e^{4A}\gamma^{-1} -\alpha,e^{4A}\gamma^{-1} -\alpha,
e^{4A}\gamma^{-1} -\alpha \right )\,.
\eea
To leading order, $e^{4A}-\alpha = \Phi_-=0$, but we will allow for 
leading order corrections of the form (\ref{KSefn}) coming from
the perturbation of $\Phi_-$. In addition, we expect other corrections
to the D3 brane potential, in particular, a mass
term for the canonically normalised radial scalar inflaton, given
to leading order by
\be
\phi = \sqrt{T_3}\; r(\eta)\,.
\ee
Note that the normalisation is strictly dependent on the 
position of the brane, which affects the volume modulus (see e.g.\
\cite{BDKKM}), and also on the trajectory of the brane, even in 
the slow roll approximation, due to the inflaton,
$\phi \leftrightarrow y^m$, being a sigma model field, \cite{sigmod}.
Putting these together, we can write the overall potential for the 
D3 brane as:
\be
T_3 V = T_3 \Phi_- + \half m_0^2 \phi^2 = T_3 \left ( \half m_0^2 \; 
[r(\eta)^2 + c_2 K(\eta) \sinh \eta \cos\vartheta]+V_0 \right )\,,
\label{D3pot}
\ee
where the constant $V_0$ is chosen so that the global minimum
of $V$ is $V=0$, and $c_2$ is an arbitrary constant.
Because $c_2$ multiplies a solution to a free Laplace equation, 
it is not fixed per se, but to keep within a self-consistent 
expansion, one would expect $c_2$ to be smaller, or of a 
similar magnitude to other terms in the potential. Indeed, it is
natural for the profile of the throat to set the scale for any 
corrections to the potential, which would mean that $c_2$ would 
be set by the deformation parameter: $c_2 \sim
\epsilon^{4/3}$. Thus the energy and pressure of the 
brane are
\bea
&& E=  T_3 \left ( e^{4A} \left[ \gamma -1 \right] +V \right )\\
&& P=  T_3 \left ( e^{4A} \left[ 1- \gamma^{-1} \right] -V\right )
\label{rhop2}
\eea

The full equations of motion (in terms of the coordinates) are:
\bea
H^2 &=& \frac{E}{3M_p^2}  \\
{\dot H} &=& -\frac{(E+P)}{2\,M_p^2}
\label{ray1}\\
\frac{1}{a^3} \frac{d}{dt}\left[ a^3 \gamma \,{\tilde g}_{mn}\, {\dot y}^n
\right]  &=& - 2\gamma\,\left(\gamma^{-1}- 1\right)^2 \, e^{4A}\, \partial_m A
+\frac{\gamma}{2} \,\frac{\partial {\tilde g}_{ln}}{\partial y^m}
\,{\dot y}^l {\dot y}^n
- \partial_m V \label{eqsfields}
\,.
\eea
The first step is to disentangle the radial and angular equations
from (\ref{eqsfields}), by a process of cross elimination. For simplicity,
we will consider motion in a single angular direction only,
writing the relevant part of the internal metric in the general form
\be
\label{gen}
ds^2= \epsilon^{4/3} \left [ \frac{d\eta^2}{6K(\eta)^2} 
+ B(\eta) d\vartheta^2 \right ]
\ee
this gives the $\eta$ and $\vartheta$ equations as:
\bea
{\ddot\eta} &=& - \frac{3 H}{\gamma^2} {\dot \eta} - 4 A' 
(\gamma^{-1}-1) {\dot \eta}^2 
- 12\epsilon^{-4/3} K^2 A'e^{4A} (\gamma^{-1}-1)^2 \nonumber \\
&& + \frac{K'}{K} {\dot \eta}^2 + 3K^2 B' {\dot\vartheta}^2
+ e^{-4A} {\dot\vartheta}{\dot\eta} \frac{V_{\vartheta}}{\gamma}
-(6K^2 \epsilon ^{-4/3}-e^{-4A}{\dot\eta}^2) \frac{V_\eta}{\gamma}
\label{etaeq}\\
{\ddot\vartheta} &=& -\frac{3 H}{\gamma^2} {\dot \vartheta} - 4 A' 
(\gamma^{-1}-1) {\dot \eta} {\dot\vartheta}
- \frac{B'}{B} {\dot \eta} {\dot\vartheta}
+e^{-4A} {\dot \eta} {\dot\vartheta} \frac{V_{\eta}}{\gamma}
-(\frac{\epsilon^{-4/3}}{B} 
- e^{-4A}{\dot\vartheta}^2) \frac{V_\vartheta}{\gamma} \;\;\;\;
\label{theteq}
\eea
These can then be solved numerically, along with the cosmological
evolution, for the relevant potential. Generally, potentials can have 
complicated angular dependence, but we confine ourselves here to 
the most simple case including only the lowest nontrivial eigenmode
with one angular direction. This should be sufficient for estimating 
the inflationary implications of brane angular dependence. 

\section{Inflationary analysis}

In order to explore the effect of angular terms in the potential, 
we use the leading order potential (\ref{D3pot}).
Because we are neither slow-rolling,
nor restricting ourselves to a specific conical region, we have to keep
the full richness of the structure of the internal space and the nonlinear
kinetic terms of the brane motion. Although the canonical inflaton field,
$\phi$, is conventionally used in expositions of slow roll inflation,
it proves more useful here to remain with the coordinate label, $\eta$,
as many of the metric functions have analytic forms in terms of $\eta$.
We focus on motion which takes place in
the angular direction, $\vartheta$, appearing in this potential,
thus from (\ref{KS6}), (\ref{gdef})
and (\ref{edef}) we can read off the function
$B = \half K \cosh \eta$, that appears in the $\vartheta$
equation of motion, (\ref{theteq}). 

Before presenting the results of the numerical analysis, a few remarks about the
various parameters are in order. First of all, the supergravity approximation
is only valid if the curvature remains large compared to the string
scale, which clearly requires the flux $g_sM \gg 1$. Secondly, as noted
by Baumann and McAllister, \cite{BMbd}, the Planck mass is
constrained by the volume of the compactification, which in
turn is bounded from the UV cut-off of the throat. Rewriting 
(\ref{Mpbd}) shows that
\be
M_p^2 > \frac{\epsilon^{4/3}g_sM^2 T_3}{6\pi} J(\eta_{UV})
\label{MpJ}
\ee
where $J(\eta)=\int I(\eta) \sinh^2 \eta$ 
is the integral in (\ref{Mpbd}), which is exponentially
growing in $\eta$ - see figure \ref{fig:Jeta}.
\FIGURE[ht]{
\epsfig{file=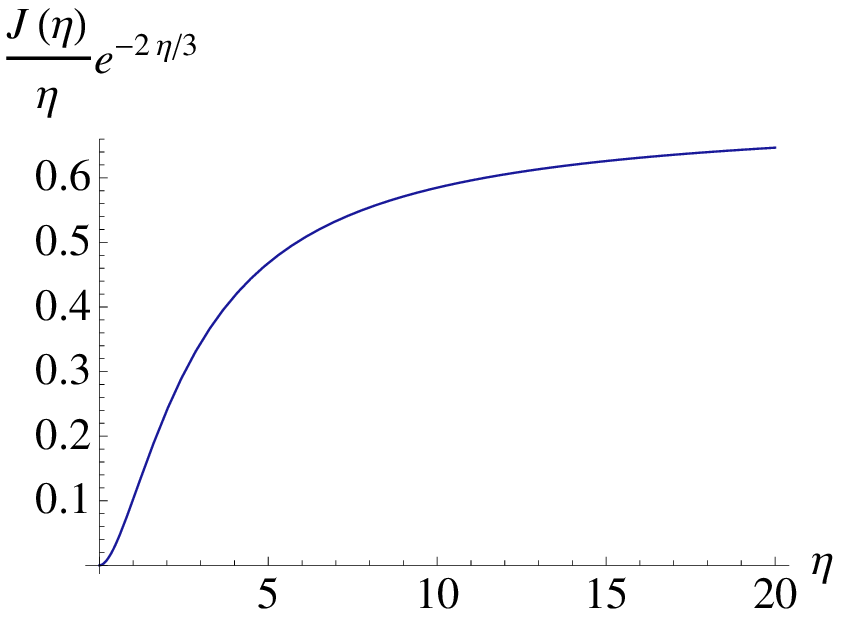,width=.6\textwidth}
\caption{Planck mass constraint: The numerical integral of
$I(\eta) \sinh^2 \eta$ with the large $\eta$ behaviour factored out.}
\label{fig:Jeta}
}

Apart from $J_{UV}=J(\eta_{UV})$, the other parameters in this
relation are set by the compactification data: $g_s M^2$ is large, 
$\epsilon$ is generally `small', and $T_3$ is determined by the 
string coupling. The hierarchy between the string scale and the
Planck scale is therefore strongly dependent on the UV cutoff 
as $J_{UV}$ grows very quickly with $\eta_{UV}$.
Generally, as the Planck mass rises, the effective scale of inflation 
is lowered, and thus the amount of inflation will drop unless 
the parameter choices allow it to persist for a correspondingly 
increased time (see the discussion after (\ref{thetaeq})).
In these models, as in slow-roll inflation, \cite{BMbd},
this bound on the Planck mass proves to be a strong constraint.
\FIGURE[ht]{
\epsfig{file=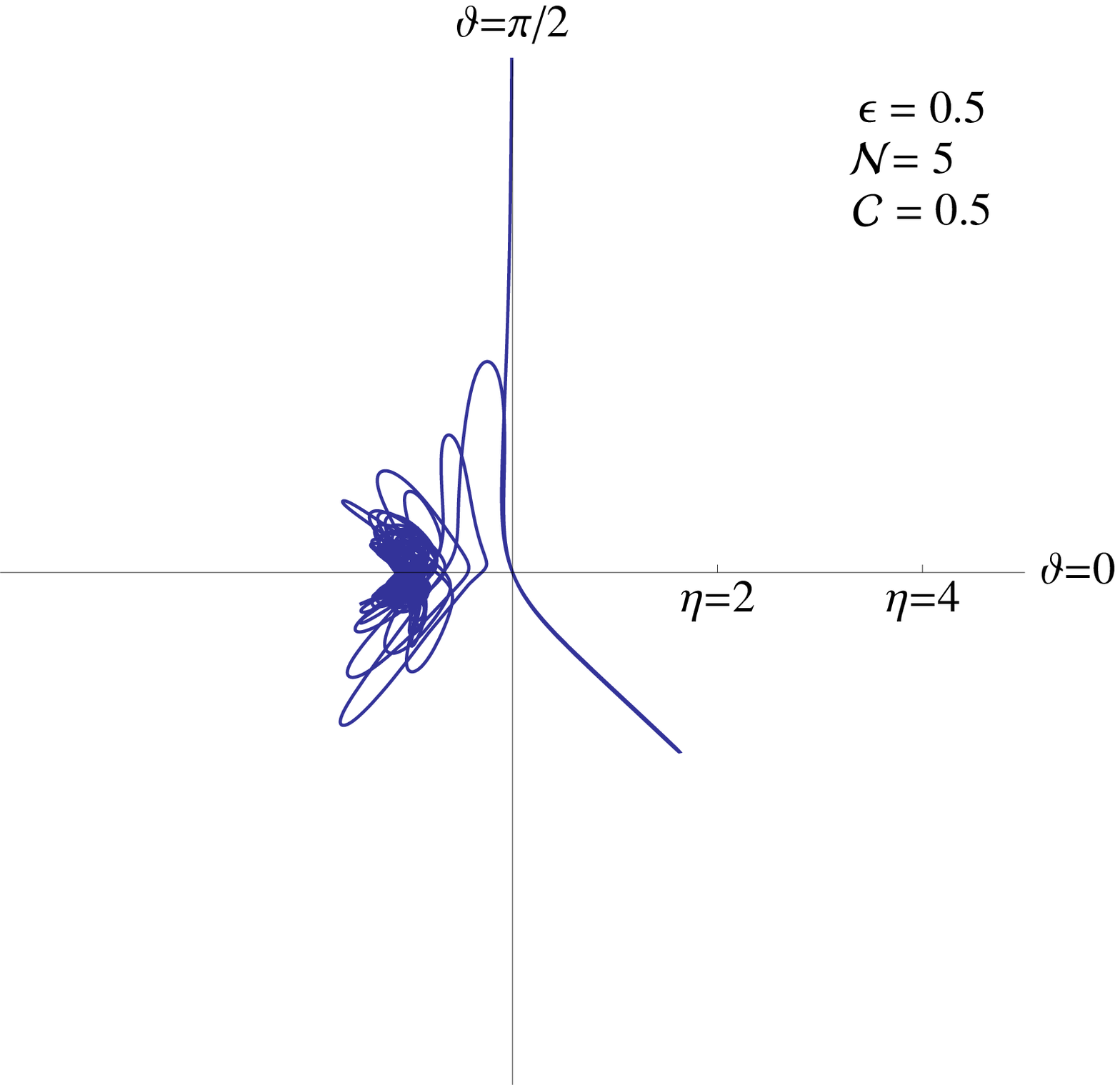,width=.45\textwidth}~~\nobreak
\epsfig{file=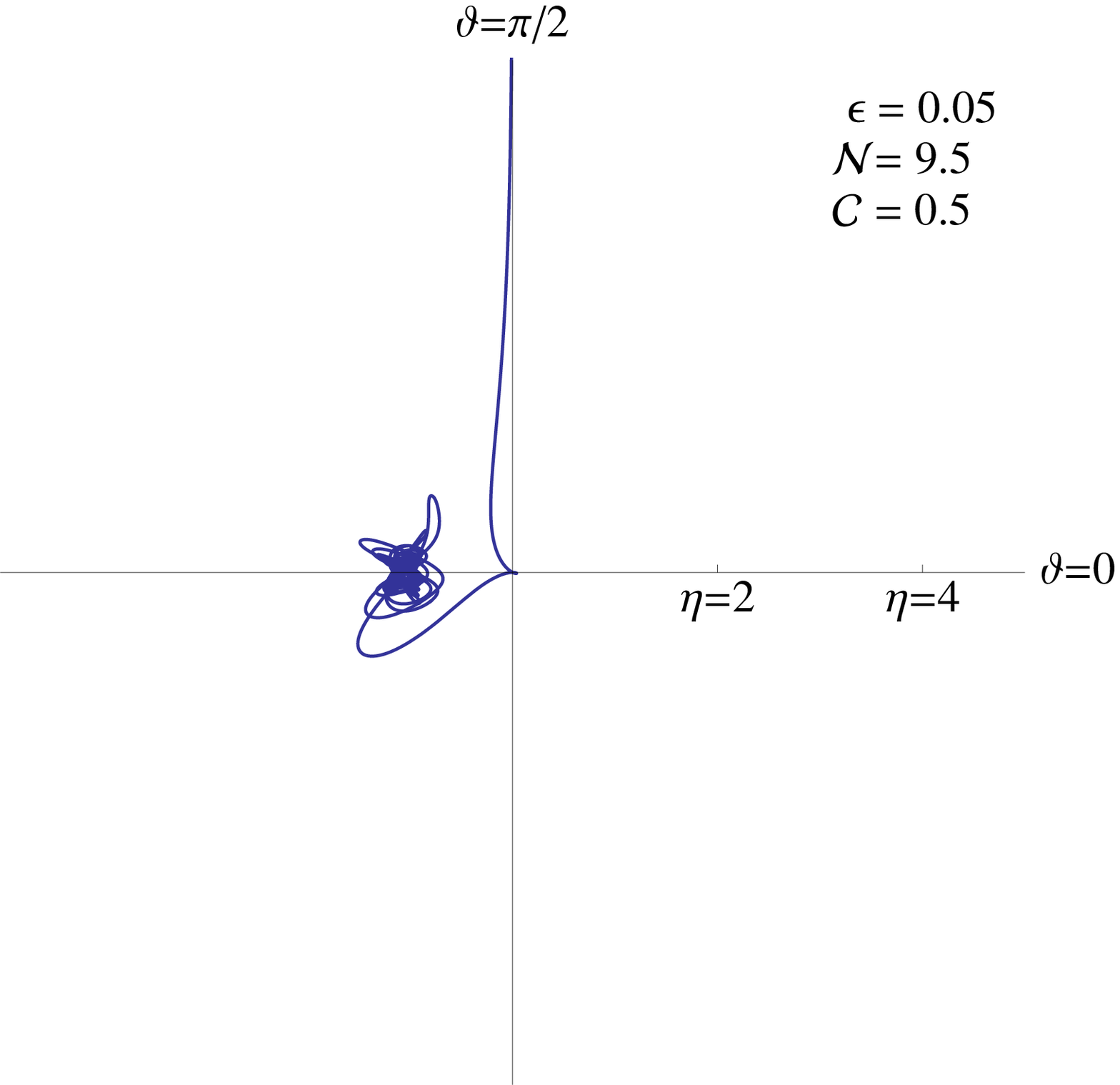,width=.45\textwidth}\\
\epsfig{file=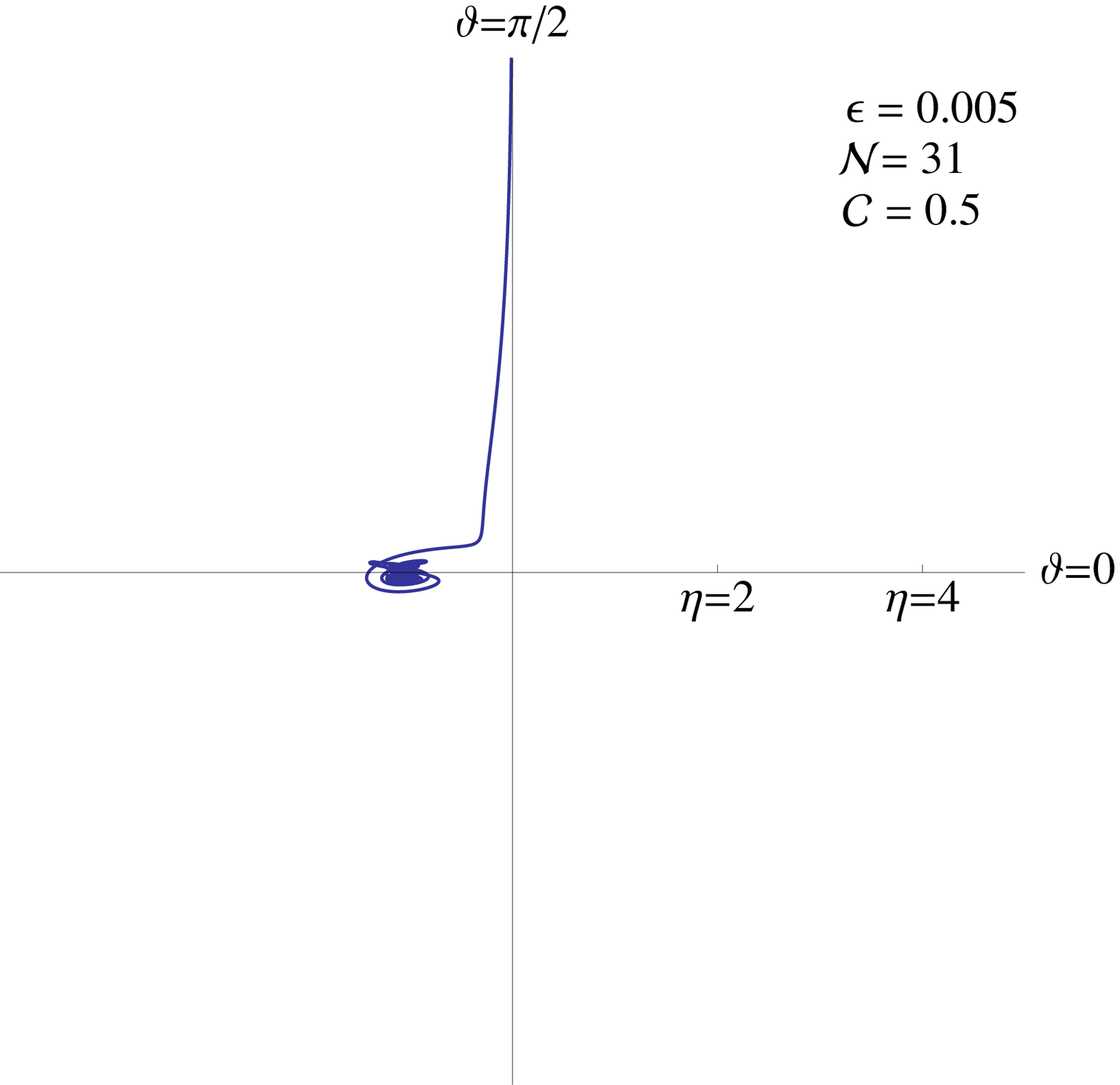,width=.45\textwidth}~~\nobreak
\epsfig{file=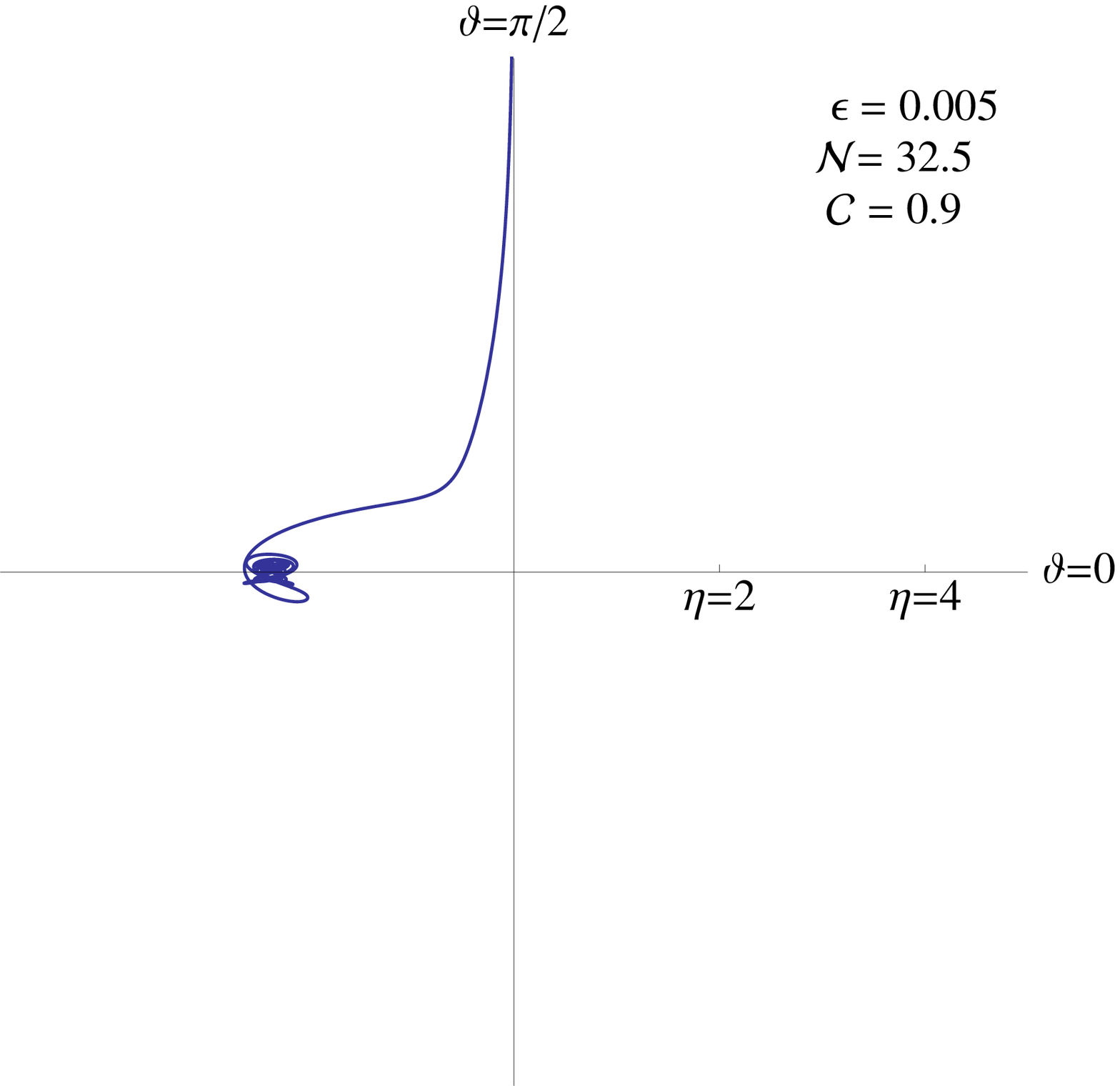,width=.45\textwidth}\\
\epsfig{file=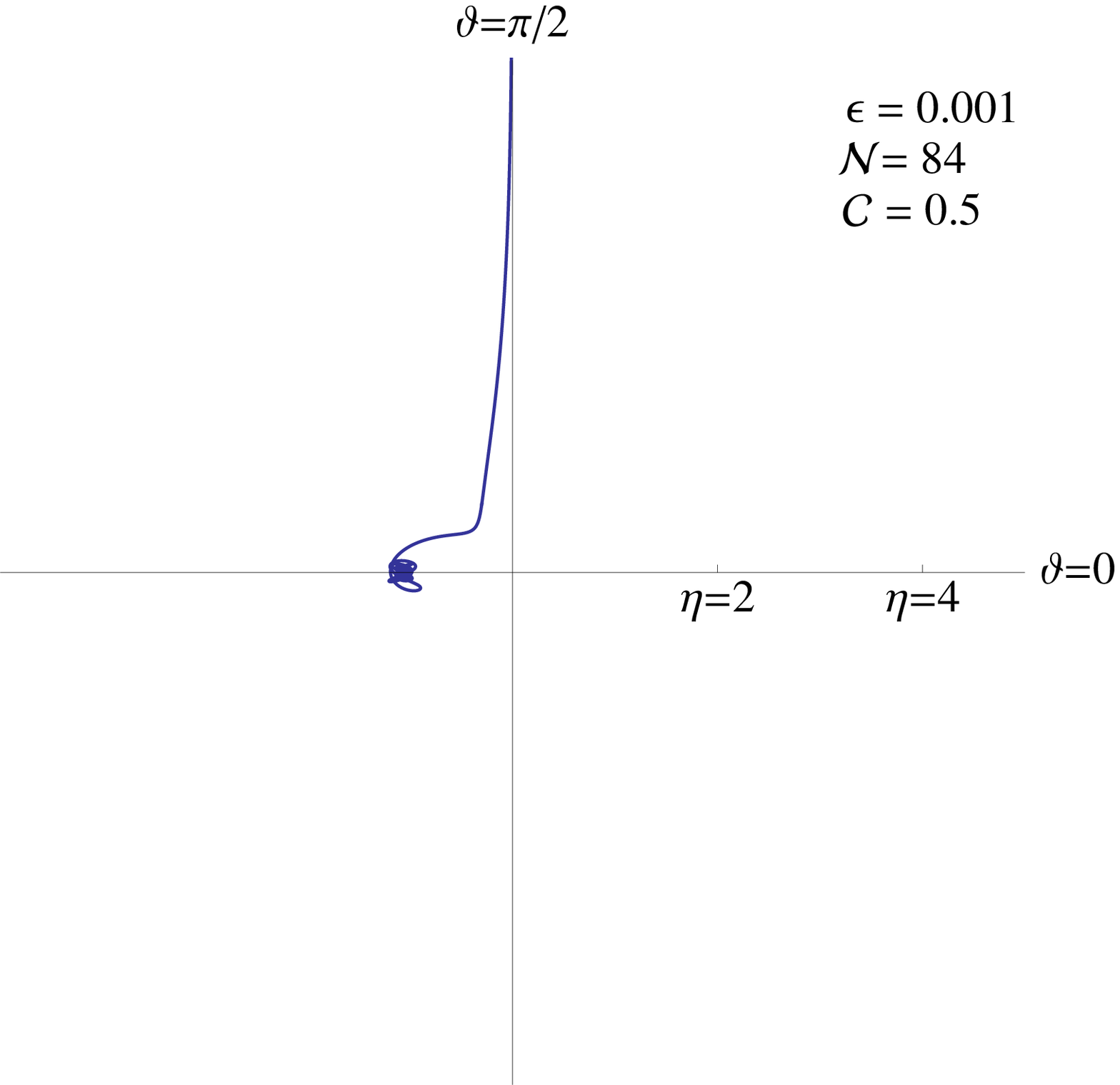,width=.45\textwidth}~~\nobreak
\epsfig{file=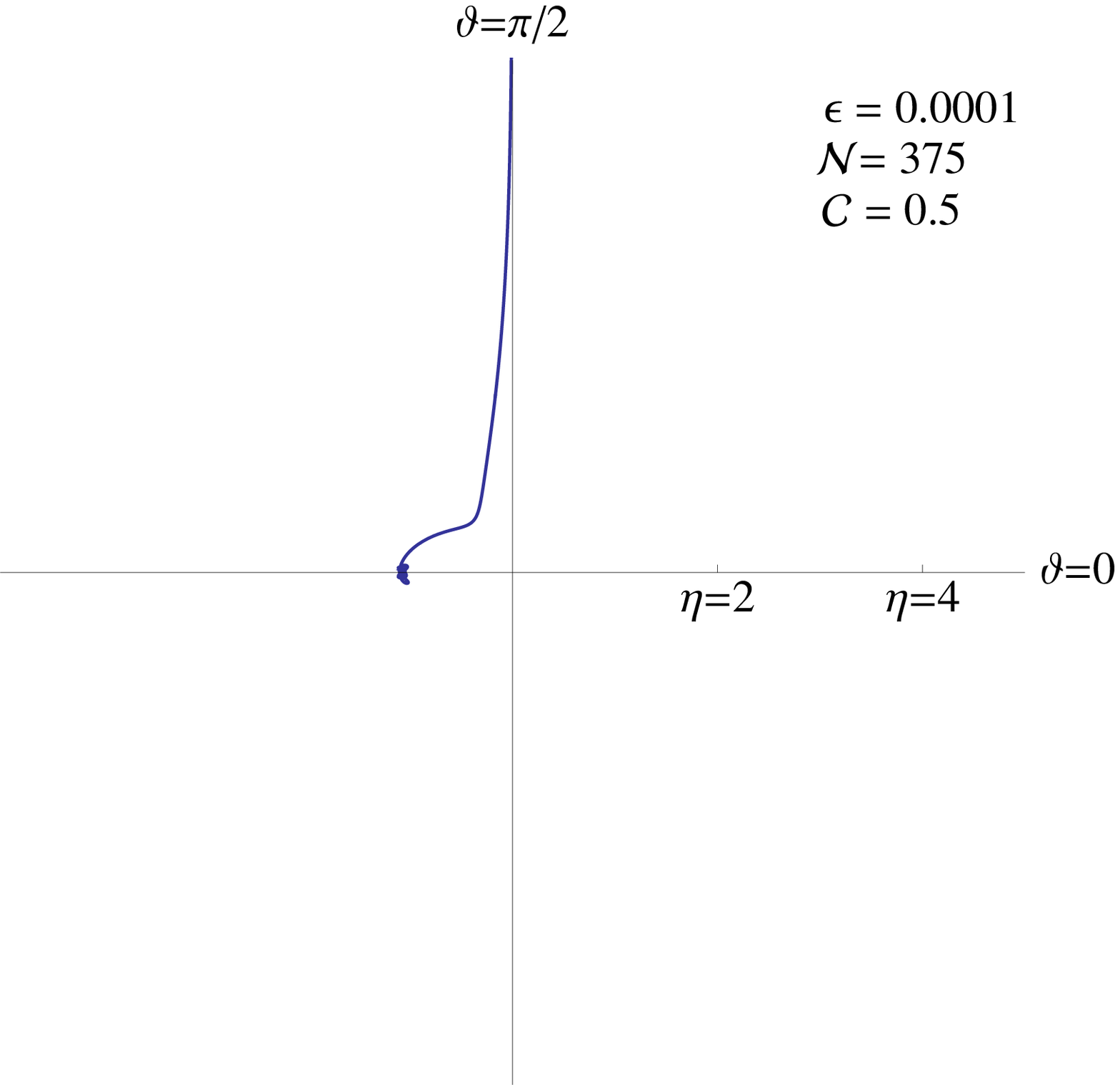,width=.45\textwidth}
\caption{Some sample inflationary trajectories with a flux 
parameter of $g_sM=100$,
inflaton mass $m_0=5$, and a saturated Planck mass. The values of 
$\epsilon$ and ${\cal C} = c_2 \epsilon^{-4/3}$, are indicated.}
\label{fig:traj}
}

To solve the full cosmological and brane equations, a numerical analysis 
is required. In order to deal with the dependence of the system on large 
(or small) parameters, we can rescale the equations of motion
by setting $\tau = \epsilon^{2/3}t/ (g_s M \alpha')$, and rewriting
the metric functions as:
\be
{\hat h} = \frac{\epsilon^{8/3}}{(g_s M \alpha')^2} 
e^{-4A} \;\;\;\; , \;\;\;\;\;\;
{\hat V} = \epsilon^{-4/3} V
\ee
so that ${\hat h}$ and ${\hat V}$ now have no explicit $\epsilon$
or $M$ dependence. Note also that
\be
\gamma^{-2} = 1 - {\hat h} \left [ \frac{{\dot\eta}^2}{6K^2} + 
B {\dot\vartheta}^2 \right ]
\ee
where a dot now denotes $d/d\tau$, also has no $\epsilon$
or $M$ dependence. 

Writing ${\hat H} = a^{-1} \frac{da}{d\tau}$, and setting $\alpha'=1$, 
the full equations of motion of the system now read:
\bea
{\hat H}^2 &=& \frac{T_3}{3M_p^2} \left [ (g_sM)^2 {\hat V} 
+ \frac{\epsilon^{4/3}}{{\hat h}} (\gamma-1) \right ] \longrightarrow
\frac{2\pi g_s}{J_{UV}} \left [ \epsilon^{-4/3} {\hat V} + \frac{\gamma-1}
{(g_sM)^2 {\hat h}} \right] \label{Hhat} \\
{\dot{\hat H}} &=& -\frac{\epsilon^{4/3}T_3}{2M_p^2{\hat h}}
( \gamma -\gamma^{-1} ) \longrightarrow -\frac{3\pi g_s 
( \gamma -\gamma^{-1} )}{(g_sM)^2 J_{UV}{\hat h}}\label{hubeq} \\
{\ddot\eta} &=& - \frac{3 {\hat H}}{\gamma^2} {\dot \eta} 
+ \frac{{\hat h}'}{\hat h} (\gamma^{-1}-1) {\dot \eta}^2 
+ 3 K^2 \frac{{\hat h}'}{{\hat h}^2} (\gamma^{-1}-1)^2 \nonumber \\
&& + \frac{K'}{K} {\dot \eta}^2 + 3K^2 B' {\dot\vartheta}^2
+ \frac{(g_s M)^2}{\epsilon^{4/3}\gamma}
\left [ {\hat h} {\dot\vartheta}{\dot\eta} {\hat V}_{\vartheta} 
-(6K^2 - {\hat h} {\dot\eta}^2) {\hat V}_\eta \right ] \\
{\ddot\vartheta} &=& -\frac{3{\hat H}}{\gamma^2} {\dot \vartheta} 
+ \left [ \frac{{\hat h}'}{{\hat h}} (\gamma^{-1}-1) 
- \frac{B'}{B} \right ] {\dot \eta} {\dot\vartheta}
+\frac{(g_s M)^2}{\epsilon^{4/3}\gamma}
\left [ {\hat h} {\dot \eta} {\dot\vartheta} {\hat V}_{\eta}
-(1 - {\hat h}B{\dot\vartheta}^2) \frac{{\hat V}_\vartheta}{B} \right ]
\hskip 8mm \label{thetaeq}
\eea
where we have shown the effect of saturating the Planck
mass bound in (\ref{Hhat}) and (\ref{hubeq}). 
We can now see the impact of the various compactification parameters.
The effect of the flux and deformation parameter is to increase the importance 
of the potential term in the brane motion, however, with the
Planck mass bound saturated, during an inflationary period with 
${\dot{\hat H}} \ll {\hat H}^2$, the inflationary scale is relatively
independent of the flux, but strongly dependent on the deformation
parameter, $\epsilon$, as well as the UV cutoff via $J_{UV}$.
Crudely therefore, we can see how increasing $m_0$ or decreasing 
$\epsilon$ will directly increase the number of e-foldings, whereas 
increasing the UV cutoff will correspondingly reduce the e-foldings 
if all other parameters are kept equal. Varying the parameter $M$ however,
would appear to have a subdominant effect, although it will alter the 
timescale of the motion via the parameter $\tau$.

For our integrations, we chose $\eta_{UV}=10$, so that
the metric functions are showing evidence of both the throat tip
deformation, as well as the asymptotic $T^{1,1}$ structure commonly 
used in the slow-roll models. We also set $2\pi g_s=1$ for convenience.
For the compactification data, we set the Planck mass at
its minimum allowed value, as determined by (\ref{MpJ}), and varied
$\epsilon$, $M$ and $m_0$, following the motion of the mobile
brane with and without the angular term in the potential. 
For initial conditions, the initial radial brane velocity was taken
to vanish, and the angular brane velocity was taken to be
highly relativistic. The initial value of the angular coordinate,
$\vartheta_0 = \pi/2$ was chosen to maximize the impact of
the angular term, when present.
For each trajectory, we followed the brane motion until it
settled into oscillations around the minimum of the potential,
counting the number of e-foldings of the associated cosmology,
to see how this varied with the model parameters.
Figure \ref{fig:traj} shows representative trajectories for the brane
as it nears the tip of the throat.

Our findings can be summarized in the following: 

\noindent$\bullet$ In terms of the trajectory
of the mobile brane, one feature that all the brane trajectories 
share is that, independent of any angular
dependence or initial momentum, the brane rapidly becomes
highly relativistic, and makes its initial sweep down the throat
in a mostly radial direction, only picking up the detail of
angular features near the tip of the throat as inflation per se ends.

\noindent$\bullet$ In agreement with the 
rescaled equations (\ref{Hhat}-\ref{thetaeq}),
varying the flux parameter slows down the brane, but makes very little
difference to the overall number of e-foldings (for the saturated Planck
mass). Increasing $m_0$, and decreasing $\epsilon$, as expected,
increases the overall amount of inflation.

\noindent$\bullet$ Increasing the angular perturbation ($c_2$) has
the effect of shifting the minimum of the potential, and thus changing the
endpoint of the brane motion, but the effect on the 
inflationary capacity of the trajectory is sub-dominant. 

\medskip

In selecting a range of sample trajectories for figure \ref{fig:traj}, 
we have focussed on varying the parameters $\epsilon$ and ${\cal C} =
c_2 \epsilon^{-4/3}$. The equations of motion (\ref{Hhat}-\ref{thetaeq})
show that varying the flux parameter slows down the brane in a similar way
to decreasing $\epsilon$, thus higher flux trajectories will look similar
to smaller deformation parameter trajectories, but with a lower number of
e-foldings. Increasing the inflaton mass however, will increase the amount
of inflation, but will not change the trajectory so dramatically. Thus, 
the main kinematical differences in the trajectories are well illustrated
by fixing the flux and inflaton mass, and varying the deformation parameter
and angular dependence.
All of the sample trajectories in figure \ref{fig:traj} show the
brane trajectory for a flux parameter of $g_sM = 100$, an inflaton
mass $m_0 = 5$, and a saturated Planck mass. 

The first two plots in figure \ref{fig:traj} show brane motion 
for relatively large values of deformation parameter: $\epsilon=0.5, 0.05$.
These show clearly how the brane moves around quite freely near the 
tip of the throat, making large oscillations before settling at the 
minimum of the potential. The brane motion is less relativistic,
however the amount of expansion can be seen to be very low (${\cal N}
\sim 5$ and $9.5$ respectively), and is not a viable inflationary scenario. 

The middle pair of plots of figure \ref{fig:traj} are more interesting.
These show the effect of changing the angular dependence of the potential,
${\cal C}=0.5, 0.9$. Here, the shift of the minimum of the potential is
quite clear. It is perhaps worth emphasizing that although at large
$\eta$, the solution of this first order Laplace equation has the same
approximate radial dependence as the inflaton mass term ($(\cosh \eta
\sinh\eta - \eta)^{1/3} \sim r^2$), at small $\eta$, this eigenfunction has
quite a different dependence, $(\cosh \eta \sinh\eta - \eta)^{1/3} \propto
\eta \propto r$, and in fact dominates over the mass term near the tip.
Although from the kinematical perspective increasing angular dependence
shifts the trajectories more significantly, the inflationary
impact of the angular terms is subdominant, as can be seen by looking at
the number of e-foldings (${\cal N} \sim 31 - 32.5$) which only
increases by about $5\%$.
Figure \ref{fig:efold} shows this in more detail by plotting the number of 
e-foldings as a function of time along an inflationary trajectory with $m_0=5$,
$g_s M =100$, $\epsilon =0.001$, comparing the amount
of inflation with and without angular
dependence in the potential. This clearly shows the subdominance
of angular terms, illustrating that the bulk of inflation occurs along
the initial, radial, sweep.
\FIGURE[ht]{
\epsfig{file=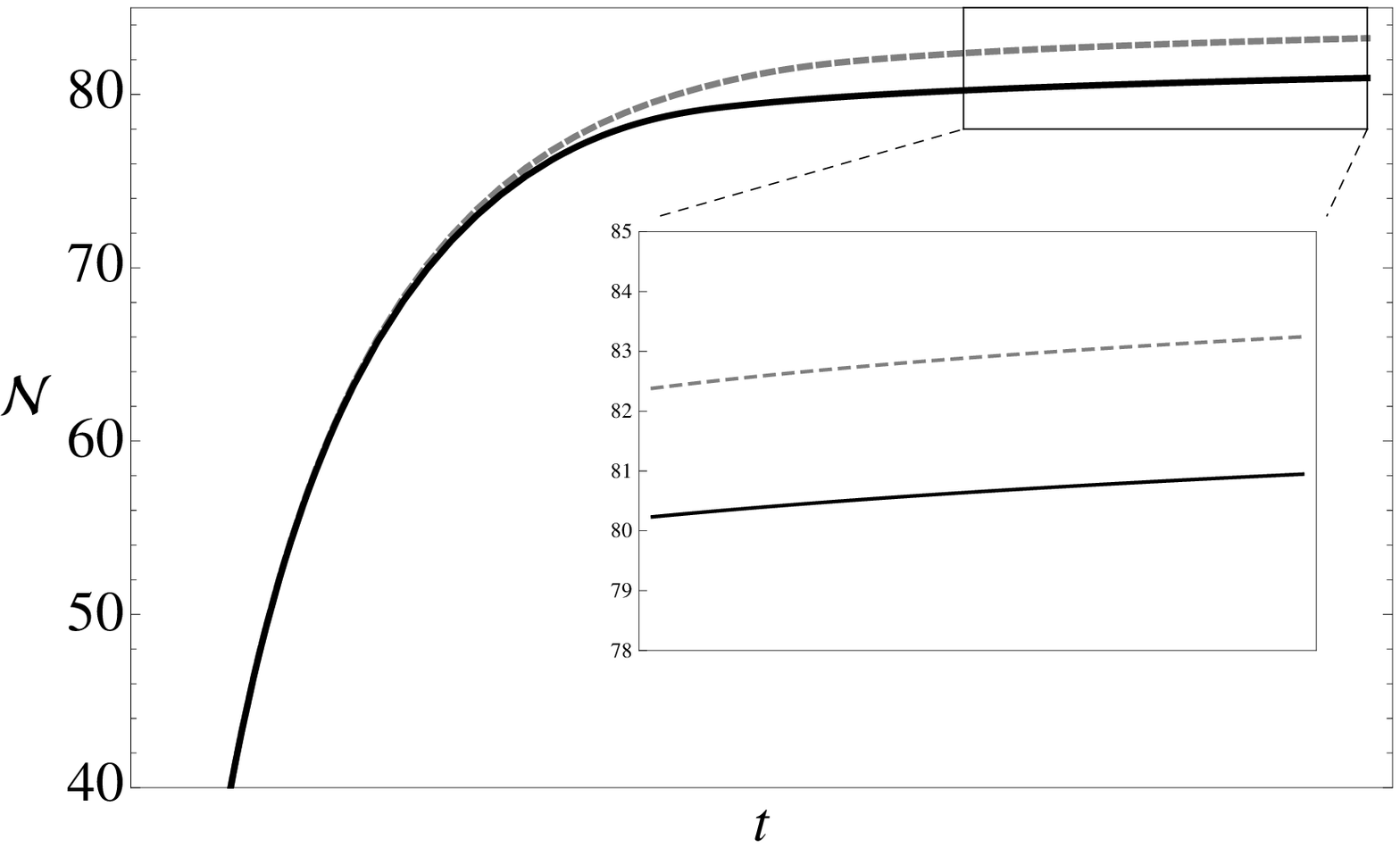,width=.7\textwidth}
\caption{The amount of inflation along a trajectory with $m_0=5$,
$g_s M =100$, $\epsilon =0.001$ with (grey dashed, $c_2 =  
\epsilon^{4/3}/2$) and without
(solid black) angular terms in the potential.}
\label{fig:efold}
}

Finally, the last two plots of figure \ref{fig:traj} show how for very 
small deformation parameters the strong warping of the throat provides a 
very strong `brake' on the coordinate speeds of the brane, giving a very
slow sweep of the brane down the throat with a large number of e-foldings.

Although these equations must be solved numerically
to extract the actual cosmological impact, an estimate for 
the number of e-foldings can be approximated analytically, which
highlights some of the dependences we have found numerically. Essentially,
we use the observation from the numerics that the bulk of cosmological
expansion occurs on the first sweep down the throat of the brane
and this motion is roughly radial. This then allows an approximate 
Hamilton-Jacobi analysis of the motion in a similar manner to that 
of the original Silversten-Tong calculation \cite{ST}.
Approximating this motion as precisely radial, 
the number of e-foldings of the universe, ${\cal N}$,
can be written as
\be
{\cal N} = \int H dt = \int \frac{H}{\dot\eta} d\eta
\ee
and we can approximate
\be
H^2 \sim \frac{T_3V}{3M_p^2} \;\;\; , \;\;\;\;
\frac{\epsilon^{4/3}e^{-4A}{\dot\eta}^2}{6K^2} \sim 1
\ee
giving
\be
{\cal N} \sim \sqrt{\frac{T_3}{3M_p^2}} \int \sqrt{\frac{V}{6}} 
e^{-2A} \frac{d\eta}{K} \leq \frac{\epsilon^{-2/3} m_0 \sqrt{4\pi g_s}}
{6J_{UV}} \;\;\; \int_{\eta_f}^{\eta_i} \frac{I(\eta) \epsilon^{-2/3} r(\eta)}{6K} d\eta
\ee
The ratio of the integral to $J_{UV}$ is explicitly independent
of the various parameters, and only depends on $\eta_{UV}$.

This argument is somewhat flawed in that it uses 
a large $\gamma$ factor to estimate the coordinate velocity, yet 
assumes that the $\gamma$ terms are
negligible compared to the potential, nonetheless, it gives a good guide as to
the general dependence of the e-foldings from brane motion near the tip, which 
is reasonably accurate for small $\epsilon$ as can be seen by comparing 
the numerical and analytical results in figure \ref{fig:Ncomp}.
\FIGURE[ht]{
\epsfig{file=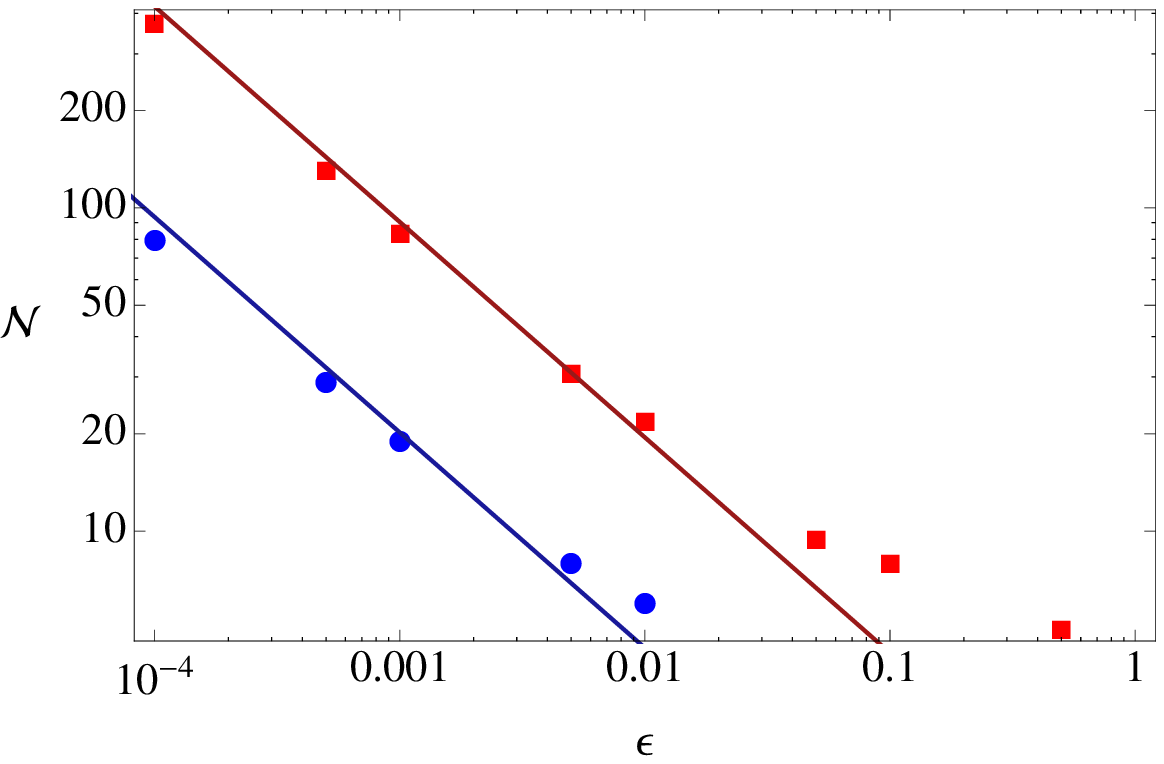,width=.7\textwidth}
\caption{The number of e-foldings as a function of the deformation 
parameter $\epsilon$ for $m_0 = 1$ (blue circles) and $m_0=5$
(red squares) with $c_2 = \epsilon^{4/3}/2$ and $g_s M = 100$. 
The analytic dependence is shown for comparison.}
\label{fig:Ncomp}
}
In this plot we see clearly that for $\epsilon \leq 0.01$, the estimate
gives a very good approximation to the dependence of ${\cal N}$ on
$\epsilon$. For strongly deformed throats ($\epsilon \geq 0.1$)
the estimate does not work so well, but since these have much
more varied angular motion and very few e-foldings, this is not
at all surprising. Unlike the small $\epsilon$ trajectories, which
spend a very long time on the initial radial sweep, exiting inflation
very close to the minimum of the potential, these trajectories fall
reasonably quickly to the tip, and cosmological expansion takes 
place equally from the initial sweep as from the following large 
angular oscillations. Finally, it also explains qualitatively the incremental 
increase in e-foldings from the angular term. Given a change in ${\cal C}$, 
the percentage increase of e-foldings is given at most by $100
(\sqrt{\frac{1+ {\cal C}_2}{1+{\cal C}_1}}-1)$, which can be seen 
to be in the ten percent range provided ${\cal C} < 1$.

\section{Summary and conclusions}

In this paper we studied generic brane motion in a warped deformed throat
incorporating full angular dependence of both the brane potential and brane
motion. This is the first such study of D3-brane inflation which samples
the many features of a warped throat, including both UV and IR features
of the geometry, as well as IR consistent supergravity corrections.
We considered a generic mass term for the D3 brane potential
as well as an analytic linearized perturbation around the ISD background,
which to our knowledge is the first closed form analytic radial 
eigenfunction on the KS background.

Our results show that angular features of the brane motion are dependent
on the compactification parameters more than on initial conditions: 
For strongly deformed throats (relatively large $\epsilon$) the 
brane `sees' the full rounded structure of the tip and has a very 
rich angular motion. These oscillatory trajectories however have very 
little cosmological expansion. Conversely, for very sharp throats 
($\epsilon\ll1$) the brane enters a strongly DBI inflating motion, 
which is geometry dominated and mainly radial in nature.
The coordinate velocities of the brane are very small, and the brane falls
to the minimum of the potential in its first sweep down the throat, 
only oscillating right at the exit of inflation and at 
the minimum of the potential.
The models and initial conditions considered were deliberately
multi dimensional, and the brane samples the full IR region of
the throat. 

In this analysis, we have deliberately considered a set-up in which
there is no obvious slow-roll regime. In particular, using the correct,
fully infrared consistent corrections to the potential, we see 
that an inflection point potential is no longer consistent. 
Our model explicitly follows brane inflation from the UV to the IR
region of the throat, and as such is not particularly suitable
for angular inflation at the tip, as the potential deep in the IR
is not sufficiently steep. However, the corrections we considered
have also been used in a restricted context in an angular tip 
inflation, \cite{tipP}, although in this case, the full dependence of
the eigenfunction on the radial direction was not considered.

In all cases, these DBI trajectories have large $\gamma$ factors and 
hence will generically generate large non-gaussianities
if used as pure inflation models in their own right, \cite{ST,NG}. 
In order to construct a viable inflationary model therefore, some
alternative mechanism must be found to produce perturbations, or this
motion must be part of a bigger inflationary picture in which CMB
scale fluctuations have already been produced (see e.g.\ \cite{tipP}).
This paper has focussed on seeking a UV/IR consistent
inflationary picture, thus we have not explored the full detail of 
cosmological perturbation theory, however, there are several options
which could, in principle be incorporated into this model. 

One particularly interesting possibility is that some curvaton 
mechanism might arise, either from one of the additional angular
degrees of freedom, \cite{DBIc} or from the vector
excitations inherent in this type of model, \cite{DWZ}.
In the curvaton picture, \cite{curv}, a (possibly more natural) option
is presented in which additional fields generate perturbations, 
rather than having the inflaton performing all roles. 

Because our angular potential is explicit, it clearly does not depend 
on several of the other internal angles, which could therefore provide 
flat directions in the potential. However, since these additional
scalar degrees of freedom are part of the multifield sigma model of
the throat, the dynamics of perturbation theory is quite involved and
only a full analysis would reveal if this is indeed possible. Perhaps
a more promising and natural approach which has recently been explored
in \cite{DWZ} is that perturbations might be generated via a vector 
curvaton, although in that case the authors did not
explore angular motion or corrections to the supergravity potential.
This would be an extremely interesting avenue to explore.

Another interesting observation is that a nonminimal gravitational 
coupling (depending on the non-radial degrees of freedom)
tends to increase the inflationary capacity, while decreasing the
$\gamma$-factor, \cite{vdb}. However, the analysis in \cite{vdb} was
from a more empirical point of view, more reminiscent of a k-inflation
model, \cite{Kinf}, and it would be interesting to see whether similar
results are attainable in an explicit supergravity model arising from
higher order corrections to the solution (e.g.\ coupling to the 
Ricci scalar) considered in this paper.

It is also possible that since the $\gamma$ factor varies throughout
the motion, for some carefully chosen parameters this problem 
could be circumvented.
It is interesting to use the intuition gleaned here to consider 
the late time evolution of more conventional slow-roll brane
models, and explore what might happen as the brane approaches 
the tip. A slow-roll model such as the delicate universe, \cite{BDKKM}, 
uses supergravity perturbations which are strictly only allowed 
in the pure $T^{1,1}$ throat,
however, setting this aside for a moment, we can speculate on the final brane
downfall to the tip. A vast scanning of parameter space for potentials
is not necessary, as the angular motion and dependence is more
a quantitative than qualitative factor, and our investigation has given
insight into the effect of the different compactification parameters as
they vary.

Typically, a slow roll model of brane inflation will be in the intermediate
adS regime of the throat geometry, and thus requires a small
deformation parameter. However, as the deformation parameter becomes 
very small, the final roll to the tip can take a long time, and be highly relativistic.
The compactification data are generally expressed differently in references
\cite{BDKKM,ABMX}, in terms of D3 flux, $N$, the warp hierarchy between
the inflating and IR region, $a_0 = e^{A(0)}$, however, knowledge 
of $M_p$, $T_3$ and the UV cut-off scale of the throat, $r_{UV}$, 
allows a translation between the different parametrizations. 
For example, to compare to the study of Agarwal et al., \cite{ABMX},  
saturating the $M_p$ bound and taking their stated parameter values
leads to deformation parameters of $\epsilon \sim 10^{-4}-10^{-7}$ 
for $a_0\sim 10^{-3}-10^{-5}$. With such small deformation parameters,
the final sweep to the throat tip could easily incorporate several to
many e-foldings of DBI inflation, and raises the question of just how much
impact this final sweep could have.

Finally, our analysis has focussed on motion in only one angular direction.
This was primarily to allow clarity of the angular terms in the potential,
but also for simplicity. We expect that including fully generic angular 
motion (which would significantly complicate the analysis)
will have an effect subdominant to the subdominant angular 
motion presented here, however, it would be useful to confirm this,
in particular to confirm or rule out the possibility of a curvaton arising
in this sector.
In addition, our analysis could be further extended by taking higher order
corrections into account, which will also have angular dependence.
While we would expect that this would be sub-dominant to the linearized
correction, it would be interesting to see precisely what the impact is
of these detailed corrections to the inflaton action.

\acknowledgments

We would like to thank Gianmassimo Tasinato for helpful discussions,
and our referee for useful comments.
This work was supported in part by STFC under the rolling 
grant ST/G000433/1. RG would like to acknowledge the Aspen Center 
for Physics, NSF grant 1066293, for hospitality while this work was
finished.

\end{document}